\documentclass[aps,preprint,onecolumn]{revtex4}
\usepackage{amssymb,amsmath}
\usepackage{graphics}
\usepackage{color}
\usepackage{hyperref}

\begin{document}
%\linenumbers

\title{Thermodynamics of tubelike flexible polymers}

\author{Thomas Vogel$^1$}
%\email{vogel@itp.uni-leipzig.de}
\author{Thomas Neuhaus$^2$}
%\email{vogel@itp.uni-leipzig.de}
\author{Michael Bachmann$^{1,3}$}
%\email{bachmann@itp.uni-leipzig.de}
\author{Wolfhard Janke$^1$}
%\email{janke@itp.uni-leipzig.de}
\affiliation{$^1$ Institut f\"ur Theoretische Physik and Centre for Theoretical Sciences (NTZ), Universit\"at Leipzig, Postfach 100\,920, 04009 Leipzig, Germany\\$^2$ J\"ulich Supercomputing Centre, Forschungszentrum J\"ulich, 52425 J\"ulich, Germany,\\$^3$ Institut f\"ur Festk\"orperforschung, Theorie II, Forschungszentrum J\"ulich, 52425 J\"ulich, Germany}

\begin{abstract}

In this work we present the general phase behavior of short
tubelike flexible polymers. The geometric thickness
constraint is implemented through the concept of the global
radius of curvature. We use sophisticated Monte Carlo
sampling methods to simulate small bead--stick polymer
models with Lennard-Jones interaction among non-bonded
monomers. We analyze energetic fluctuations and structural
quantities to classify conformational pseudophases. We find
that the tube thickness influences the thermodynamic
behavior of simple tubelike polymers significantly, i.e.,
for given temperature, the formation of secondary structures
strongly depends on the tube thickness.

\end{abstract}

\maketitle

\section{\label{sec:intro}Introduction}

To resolve the conformational mechanics of
secondary-structure formation is one of the major tasks in
polymer science. While in the ``real world'' experiments are
restricted to specific molecules under specific conditions,
in the ``virtual world'' of computer simulations there are
no such limitations. Using reasonably simplified models,
systematic studies of classes of polymers in different
environments are possible~\cite{binder95buch}.

A common, effective and widely-used coarse-grained model for
polymers is the bead--stick
model~\cite{flory53buch,degennes_book}. Here the polymer is
modeled as a linear chain of pointlike monomers, which
correspond to molecular units, e.g., amino acid residues in
the case of proteins. The monomers are connected by stiff
bonds and interact via simple effective pair
potentials. This class of models enables computer
simulations of very large polymer systems and is,
for example, quite useful for studying global structure
formation or structural
transitions~\cite{grassb97pre,voge07pre,schna09cpl}.  On
the other hand, due to the simple pairwise interactions, it
is hardly possible to investigate the formation of secondary
structures in a systematic way, which is due to missing
specific extensions like hydrogen bonds, anisotropy,
explicit stiffness,
etc.~\cite{noguch98jcp,kemp98prl,rapa02pre,sabeur08pre}.

In this work, we therefore follow the approach introduced by
Banavar and Maritan \textit{et al.}~\cite{banamarit03jpcm,banagonz03jsp,bana04pre}, where a
tubelike model is considered instead of linelike chains. The
virtual thickness of the tube caused by the bulky shape of
the monomers (e.g., because of side chains connected to the
backbone) is introduced via a three-body interaction. 
The general tertiary phase behavior of tubelike polymers with 40 and more 
monomers has already been investigated using a square-well
model~\cite{banamarit03jpcm}, identifying the folding and
collapse transitions in a structural phase diagram
parameterized by thickness and temperature. In our study, we
investigate in detail the thickness and temperature
dependence of secondary-structure formation of tube
polymers, employing a continuum model with inter-monomeric
Lennard-Jones potential. For this reason, we consciously
investigate rather small chains (with up to 13
monomers). For longer chains, tertiary folding effects
become apparently important and symmetry, anisotropy, and
marginal compactness of globular protein structures are then
doubtlessly of interest~\cite{hoang1,bana07arbbs}.  However, the
globular arrangement of secondary segments in tertiary folds
is not in the focus of this study and it is also hardly
feasible to perform a similarly precise analysis of the
present work for longer chains.

The present work extends our
recent study of ground-state properties of tubelike polymers
as a function of their thickness~\cite{letter,partI}. 
Thus, the conformations identified in these former studies 
represent the dominant structures in the 
fluctuation-free, i.e., lowest-temperature region ($T\to 0$) of the entire conformational 
phase diagram that we will discuss in detail in the following.
Thus, the goal of this study is to identify independently of the chain length the relevant
pseudophases in the thermodynamic phase diagram, based on the shape of
ground-state structures.
The notion ``phase'' shall be handled with some care; conformational phase
transitions of small systems are not thermodynamic phase transitions in a strict
sense. Nonetheless, there is a strong similarity of these structural transitions and thermodynamic
phase transitions, as both are typically governed by the competition of energy and entropy.
However, to make clear that there can also be significant differences (no collapse
of fluctuating quantities, i.e., there are transition regions rather than transition 
points), we call conformational phases of short chains ``pseudophases'' in the following~\cite{bj1}.

Recent related studies also apply other tube models for
homopolymers to investigate secondary-structure formation
(see, e.g.,
Refs.~\cite{hoang04pnas,snir05sci,snir07pre,auer07prl,wolff08gene}).
These are, however, based on different approaches to influence
or potentiate structure formation. See, e.g., the discussion
in Ref.~\cite{letter}.

The structure of the rest of the paper is as follows: In
Sec.~\ref{sec:models}, we describe the model and specify the
simulation methods we employed. In
Sec.~\ref{sec:results_homo}, we present the complete
thermodynamic phase diagrams for various chain lengths of
homopolymers and analyze and classify the different
pseudophases therein. In Sec.~\ref{sec:AB}, we introduce a
hydrophobic-polar heteropolymer tube model and analyze the
ensuing pseudophase behavior for an exemplified sequence of
monomers. Finally, our main findings are summarized in
Sec.~\ref{sec:summary}.

\section{Model and Methods}
\label{sec:models}

As outlined above, we employ in this study a linear,
flexible polymer model with thickness, i.e., we consider
tubelike chains instead of linelike objects.
The bond length in this model is kept fixed, i.e.,
$r_{i,i+1}=1$, where $r_{i,j}=|\,\mathbf{x}_i-\mathbf{x}_j|$
denotes the distance between two monomers. The monomers
interact via a standard Lennard-Jones (12,6)-potential
resulting from pairwise attractive van-der-Waals and
short-range repulsion forces:
\begin{equation}
V_{\mathrm{LJ}}(r_{i,j})=
4\epsilon\left[\left(\frac{\sigma}{r_{i,j}}\right)^{12}
-\left(\frac{\sigma}{r_{i,j}}\right)^6\right]\,.
\label{eq:LJ}
\end{equation}
In the following, we set $\sigma=\epsilon=1$, such
that $V_{\text{LJ}}$ vanishes for $r_{i,j}=1$ and is minimal
at $r_{i,j}^{\text{min}}=2^{1/6}\approx1.122$ with
$V_{\text{LJ}}(r_{i,j}^{\text{min}})=-1$. The total energy of
a conformation
$\mathbf{X}=(\mathbf{x}_1,\ldots,\mathbf{x}_N)$ is then
calculated as the sum of all LJ contributions,
$E(\mathbf{X})=\sum_{i,j>i+1}V_{\text{LJ}}(r_{i,j})$.

\begin{figure}
\includegraphics{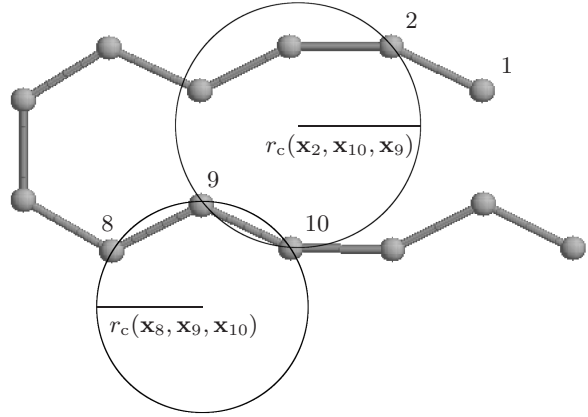}
\caption{Two examples of circumcircles of three monomers and the
corresponding radii of curvature. The small circle
corresponds to the radius of curvature of three consecutive
monomers, i.e., to the local radius of curvature
of the monomers $(i,i+1,i+2)=(8,9,10)$. The
bigger circle corresponds to the radius of curvature
$r_\text{c}$ of the monomers $(i,j,k)=(2,10,9)$.}
\label{fig_grc}
\end{figure}

To define the thickness of a conformation $\mathbf{X}$, we
apply the concept of the global radius of
curvature~\cite{gonzmad99pnas}. Accordingly, we measure all
(see technical remark below) radii of curvature
$r_\text{c}(\textbf{x}_i,\textbf{x}_j,\textbf{x}_k)$, i.e.,
the radii of the circles defined by the monomer positions
$\textbf{x}_i$, $\textbf{x}_j$, and $\textbf{x}_k$.  The
minimal radius of curvature is called the global radius of
curvature:
\begin{equation}
r_\text{gc}(\mathbf{X})
\mathrel{\mathop{:}}=
\min\{r_\text{c}(\textbf{x}_i,\textbf{x}_j,\textbf{x}_k)\,|\,
\forall\,i,j,k;i\neq j\neq k\}\;.
\label{eq:grc}
\end{equation}
The thickness $d(\mathbf{X})$ of the polymer tube is simply
twice the global radius of curvature,
$d(\mathbf{X})=2\,r_\text{gc}(\mathbf{X})$.  For
illustration, Fig.~\ref{fig_grc} shows two radii of
curvature of a conformation with $N=13$ monomers. As a
technical remark: The explicit calculation of all radii of
curvature is obviously needless and would be very
expensive in terms of computing time as the number of radii
grows with the third power of the monomer number
($\mathcal{O}(N^3)$). By excluding a huge number of a priori
too large radii with much less effort, the calculation can
be done nearly in $\mathcal{O}(N\log N)$ steps (possibly
plus some marginal higher-order terms)~\cite{neuh06}.
In order to simulate the model, we restrict the
conformational space by a thickness constraint $\rho$, such
that conformations with $r_\text{gc}<\rho$ are forbidden,
i.e., the Heaviside function
$\Theta(r_\text{gc}(\mathbf{X})-\rho)$ is included in the
partition function,
\begin{equation}
Z=\int\mathcal{D}X\,\Theta(r_\text{gc}(\mathbf{X})
-\rho)\,\text{e}^{-\beta E(\mathbf{X})}\,,
\end{equation}
where $\mathcal{D}X$ is the functional integral measure and
$\beta=1/k_\text{B}T$ is the inverse temperature (with
$k_\text{B}=1$ in natural units). For a more detailed
description and discussion of the concept and its
applicability to polymer models, see, for example,
Refs.~\cite{banagonz03jsp,bana04pre,letter,partI}.

In the Monte Carlo simulations, we use multicanonical flat
histogram sampling~\cite{bergneuh91plb,bergneuh92prl} to
estimate the density of states. To determine the weight
factors, we employ the recursive method of Wang and
Landau~\cite{wangl01prl}, with the control parameter $f$
initialized and subsequently decreased to $f-1<10^{-7}$ as
described in Ref.~\cite{zhoubhatt05pre}.
We remark that for any finite value
$f-1$ the Markov chain of configurations, as generated with
the Wang--Landau algorithm does not possess a proper Gibbs
measure. Rather, the density of states, entering here the
Metropolis criterion, is constantly updated and hence varies
as the Markov chain proceeds. Thus detailed balance is
violated in particular in the initial simulation phase. We
therefore decided to freeze the weights at some point of the
Wang--Landau iteration and to perform a multicanonical
production run with a Gibbs measure as determined by the
multicanonical weight factor. Furthermore, we also checked
our results for reliability against data obtained by
parallel tempering
simulations~\cite{partemp1,partemp2,neuh06,bittner08prl},
which generate simultaneous ensembles of polymers at a
multitude of temperature values. The checks are done for
selected parameter sets, as well as against data from the
study presented in Ref.~\cite{arkin05pre}. The simulations
of different polymer lengths and thickness constraint values
were carried out separately to avoid correlations and
statistical \hbox{imbalances}.

\section{Conformational phase diagrams\\ of tubelike homopolymers}
\label{sec:results_homo}

\subsection{General}

\begin{figure*}
\includegraphics{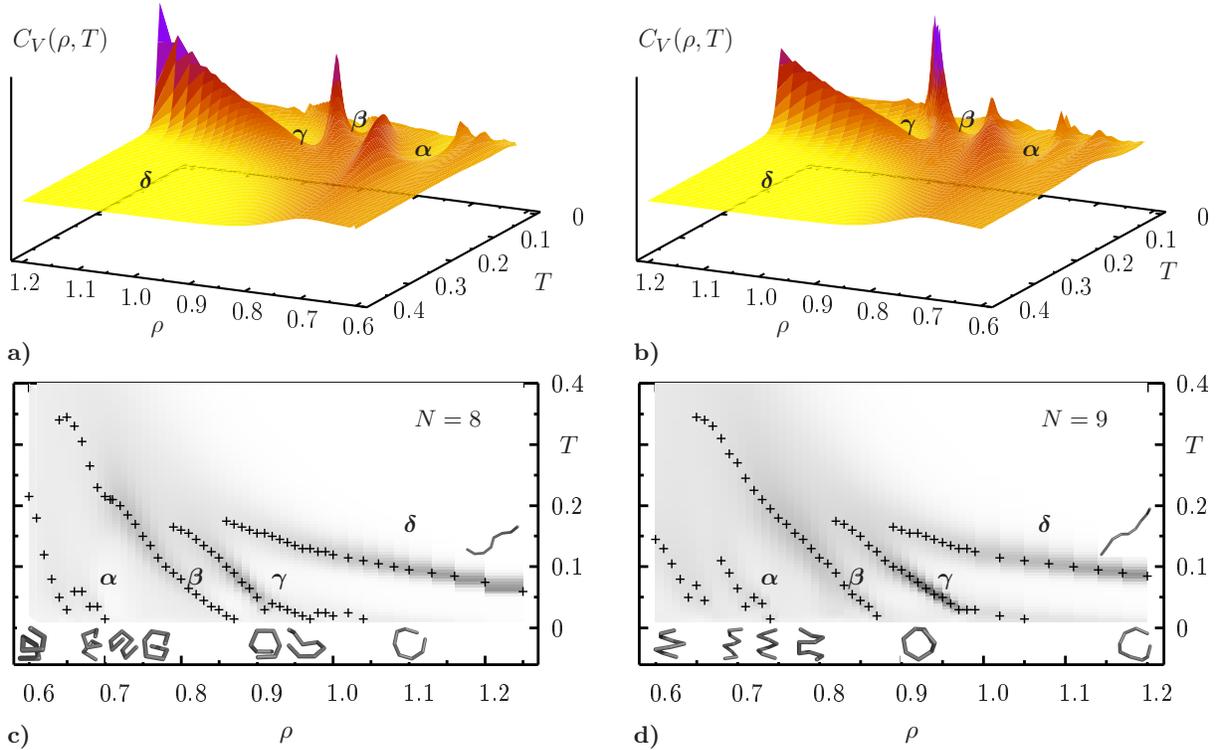}
\caption{(color online) Phase diagrams of the homopolymers with $N=8$
(left) and $N=9$ (right). The labels
$\alpha$, $\beta$, $\gamma$, and $\delta$ indicate the
different pseudophases at finite temperature. Figures \textbf{a)} and \textbf{b)} show the perspective view of the
specific-heat landscape, and in \textbf{c)} and \textbf{d)}, the
top-views are plotted with marked peak positions for various
parameters $\rho$. The specific-heat values are encoded in
gray scale. The pictures in the insets in
\textbf{c)} and \textbf{d)} correspond to the ground states
presented in Ref.~\cite{partI}, the pictures in the $\delta$
regions show relevant conformations there.}
\label{fig1}
\end{figure*}
\begin{figure*}
\includegraphics{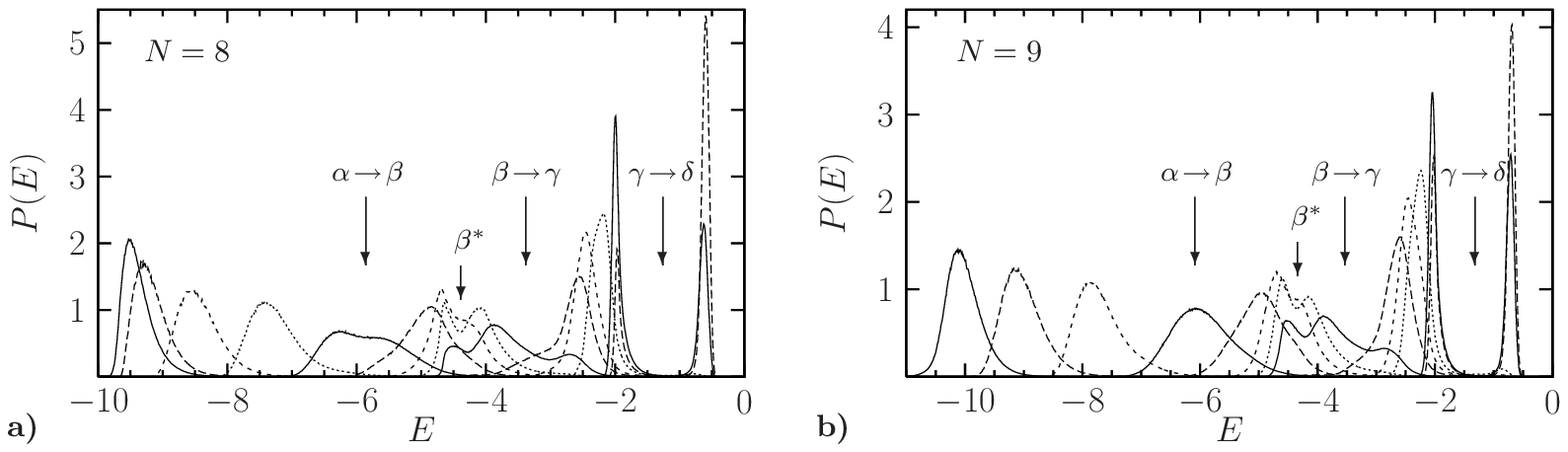}
\caption{Energy histograms for various thickness constraints $\rho$
at $T=0.1$. Histograms corresponding to specific-heat maxima
are marked with arrows. \textbf{a)} $N=8$
polymer. Histograms correspond to the following thickness
parameters: $\rho=0.7$ (solid line), $0.72$,
$0.74$, $0.76$, $0.78$ (solid line,
$\alpha\to\beta$), $0.8$, $0.82$, $0.84$, $0.86$
(solid line, $\beta\to\gamma$), $0.88$, $0.9$,
$0.95$, $1.08$ (solid line, $\gamma\to\delta$),
$1.13$.
\textbf{b)} $N=9$
polymer. Histograms correspond to $\rho=0.72$ (solid
line), $0.75$, $0.78$, $0.81$ (solid line,
$\alpha\to\beta$), $0.83$, $0.85$, $0.87$, $0.89$
(solid line, $\beta\to\gamma$), $0.92$, $0.95$,
$1.11$ (solid line, $\gamma\to\delta$), $1.14$.
The histograms were obtained by reweighting the density of
states and are consistent with histograms obtained from
independent canonical simulations at this temperature. These
histograms contain about $10^{10}$ entries. Statistical
errors are less than 1\% and, almost everywhere, smaller
than the line width.}
\label{fig8IV-Vb}
\end{figure*}
In this work, we study homopolymers consisting of $N=8$,
$9$, $10$, and $13$ monomers. After having considered the
low-temperature regime, i.e., ground states, in a recent
paper~\cite{partI}, we here concentrate on the
conformational phase behavior at finite temperatures. As
common, we calculate the specific heat and consider the peak
regions of this observable as indicators of relevant
thermodynamical activity.  Figure~\ref{fig1} shows these
specific-heat landscapes for the $N=8$ and $N=9$
polymer. The points (+) plotted in the top-view
representation of Fig.~\ref{fig1}\,c)
resp. Fig.~\ref{fig1}\,d) indicate the positions of the
crest lines in this landscape, i.e., the lines signaling
structural changes. We notice four major pseudophases, which
we denote by $\alpha$, $\beta$, $\gamma$, and~$\delta$.
In Fig.~\ref{fig8IV-Vb}, we show the corresponding canonical
energy histograms at temperature $T=0.1$ for different
thickness constraints $\rho$, with the histograms at the
transition values of $\rho$ marked by
arrows. Both plots, for $N=8$ and $N=9$, do not
differ qualitatively, i.e., have all interesting features in
common. The phase structure will be discussed in the
subsequent detailed analysis of the pseudophase diagrams.

In the insets of Figs.~\ref{fig1}\,c) and~\ref{fig1}\,d),
ground-state conformations, according to their thickness,
are shown. They provide a first indication for the
population of the respective pseudophase at finite
temperatures. Deeper analyses will strengthen the
expectation that the ground-state conformations are the
relevant conformations in the corresponding pseudophases at
finite temperatures as well. This includes, for example, the
analyses of distributions of structural observables like
end-to-end distance, radius of gyration, radial distribution
of monomers, bond angles and torsion angles, as well as
comparisons with reference structures and ``counting''
structural components, e.g., using pattern
recognition~\cite{tenenbaum00science}, during additional
canonical simulations at fixed temperatures. Let us note,
that we neglect data for $\rho\lesssim0.6$, which
corresponds to the pure Lennard-Jones volume exclusion, as
the thickness constraint does not influence the system at
all below this ``natural thickness''~\cite{partI}.

\subsection{Analysis of structural phases}

\begin{figure*}
\includegraphics{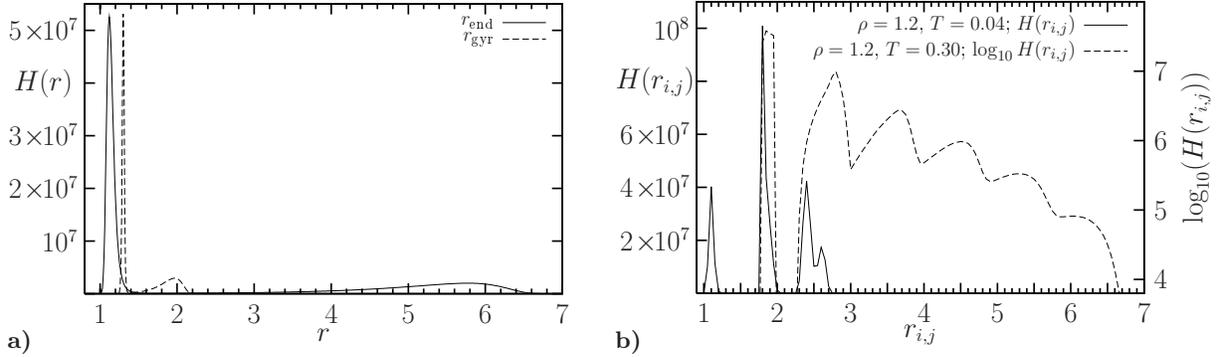}
\caption{Measured histograms from simulations at fixed temperatures
for the $N=8$ polymer \textbf{a)} at the transition from the
bended-ring phase $\gamma$ to the sprawled-coil phase
$\delta$ and \textbf{b)} deep inside these two
phases. \textbf{a)} End-to-end distance (solid
line) and radius of gyration (dashed line) at the
$\gamma\leftrightarrow\delta$ transition ($\rho=1.08$ and
$T=0.1$). \textbf{b)}~Radial distribution ($\rho=1.2$) in
the bended-ring phase $\gamma$ (solid line,
$T=0.04$) and in the sprawled-coil phase $\delta$
(dashed line, $T=0.3$). The histograms are
differently scaled for better visibility, each contains more
than $10^{9}$ entries. Statistical errors are less than 1\%
and smaller than the line width.}
\label{fig8IV-Va}
\end{figure*}
We begin the detailed discussion of the different structural
phases with the high-thickness region, i.e., with the phase
$\gamma$ and the transition to $\delta$. Based on the
knowledge of the ground states and some general structural
properties of polymers, we assume in $\gamma$ a population
of bended rings, which undergo a structural change to
sprawled random coils in $\delta$, which become more and
more rodlike with increasing thickness. This assumption can
be illustrated and strengthened by an example in little more
detail. For $N=8$ monomers, let us consider the geometrical
objects ``octagon'' and ``straight line'' as limiting
prototypes of these regions. Calculating the properties of
these prototypes, one expects for the end-to-end distance
distributions a sharp peak at the position of the LJ
potential minimum, i.e., at $r_\text{end}\approx1.12$, and a
diffuse peak at $r<7$, for the radius of gyration
distribution a sharp peak at $r_\text{gyr}\approx1.3$ and a
diffuse peak at $r<2.34$, and for the radial distribution
function sharp peaks at $r\approx1.1$, $1.8$, $2.35$, and
$2.55$ and smooth peaks below integer values for the
respective conformations. In Fig.~\ref{fig8IV-Va}, these
distributions are shown, measured in canonical
simulations at the transition temperature and within both phases. In
Fig.~\ref{fig8IV-Va}\,a), the end-to-end distance and radius
of gyration histogram are plotted, and Fig.~\ref{fig8IV-Va}\,b)
shows the radial distribution function. These quantities
exhibit exactly the assumed behavior, i.e., the peaks of the
measured distributions appear exactly at the calculated
values for the anticipated ``prototypes''.
Additionally, the bimodal shapes of the distributions in
Fig.~\ref{fig8IV-Vb} at the transition $\gamma\to\delta$ are
an indication for the first-order-like character of the
transition with coexisting conformational phases. The energy
histograms near the transition point exhibit two distinct
peaks separated by broad energy gaps. During simulations at
the transition line, both structures appear equally, as can
be seen for example in~Fig.~\ref{fig8IV-Va}\,a).

\begin{figure*}
\includegraphics{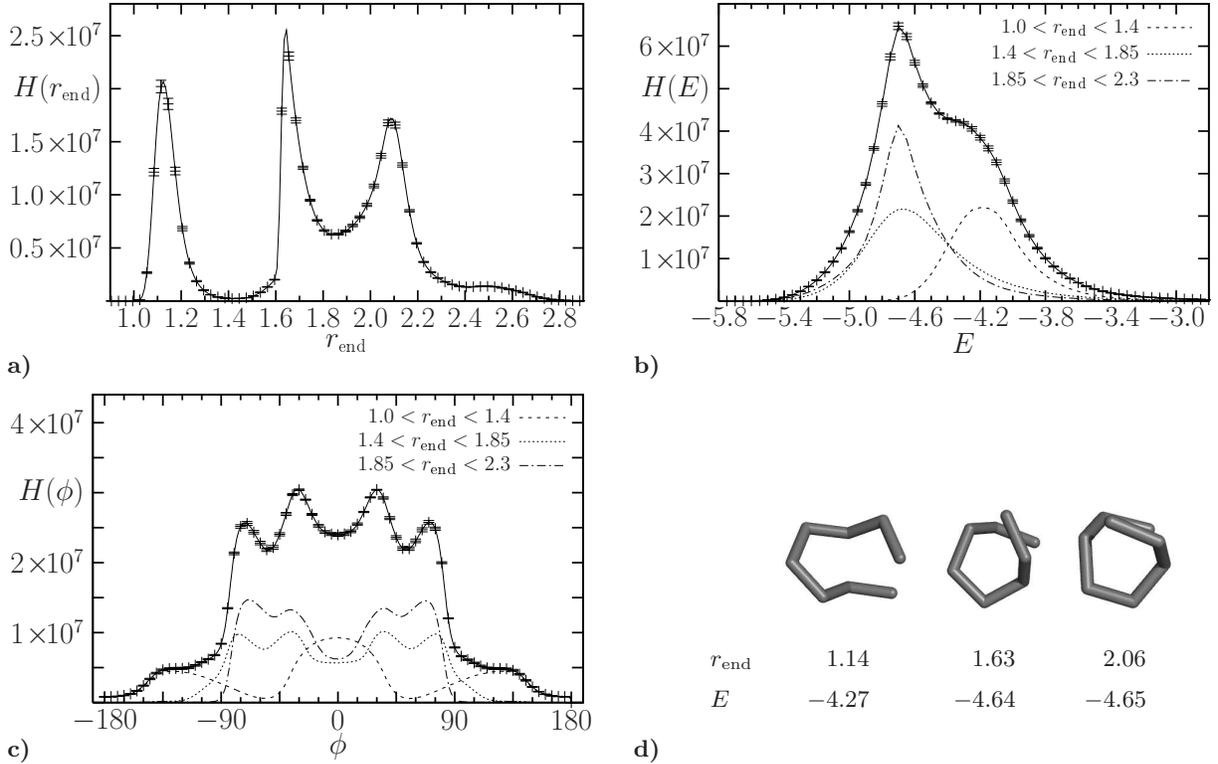}
\caption{Measured histograms in phase $\beta$ for
$\rho=0.82$ and $T=0.1$ for the $N=8$ polymer. \textbf{a)}
The end-to-end distance histogram exhibiting three separate
peaks indicating three different major contributing groups of
conformations. \textbf{b)} The energy histogram and
\textbf{c)} the histogram of torsional angles. Error bars were
obtained from independent simulations and are shown
exemplarily. In \textbf{b)} and \textbf{c)}, the histograms
for each group of conformations, distinguished by its
end-to-end distance, are shown in addition. Each histogram
contains at least $10^9$ entries. \textbf{d)}~Representatives
of each group of this energetic pseudophase and their
corresponding properties.}
\label{fig8III}
\end{figure*}
Reducing the thickness parameter $\rho$, we reach the phase
$\beta$, which we call the sheet phase. Figure~\ref{fig8III}
shows the results of simulations at $\rho=0.82$ and $T=0.1$
for the $N=8$ polymer, which belongs to the region called
$\beta^*$ in Fig.~\ref{fig8IV-Vb}\,a). There are mainly
three structures dominating the phase $\beta$, amongst them
the two ground-state conformations in the range
$0.89\leq\rho\leq0.99$ (cp. Ref.~\cite{partI} and
Fig.~\ref{fig1}). As shown in Fig.~\ref{fig8III}\,a), they
can be distinguished with the help of the end-to-end
distance, where three distinct peaks in the distribution
appear, whereas they cannot be resolved in terms of the
specific heat. The plot in Fig.~\ref{fig8III}\,b) shows the
overall energy distribution as well as the contributions
from the three regions corresponding to the peaks in the
end-to-end distribution. As illustrated in
Fig.~\ref{fig8III}\,d), the peak in the energy distribution
is associated with ring-like conformations and their
excitations, whereas the shoulder is caused by hairpin-like
conformations. In Fig.~\ref{fig8III}\,c), we plot the
distribution of torsion angles. The contributions of the
different structural classes can be distinguished very well
again. One notes for example an accumulation of torsion
angles around $\phi=0$ in the contribution of the
hairpin-like conformations, an indication for the planar
structure of the conformation. At $\beta^*$, the
conformations extend into the third dimension, i.e., bonds
within the conformations begin to overlap. An analogous
behavior is found for $N=9$, see Fig.~\ref{fig8IV-Vb}\,b).

\begin{figure*}
\includegraphics{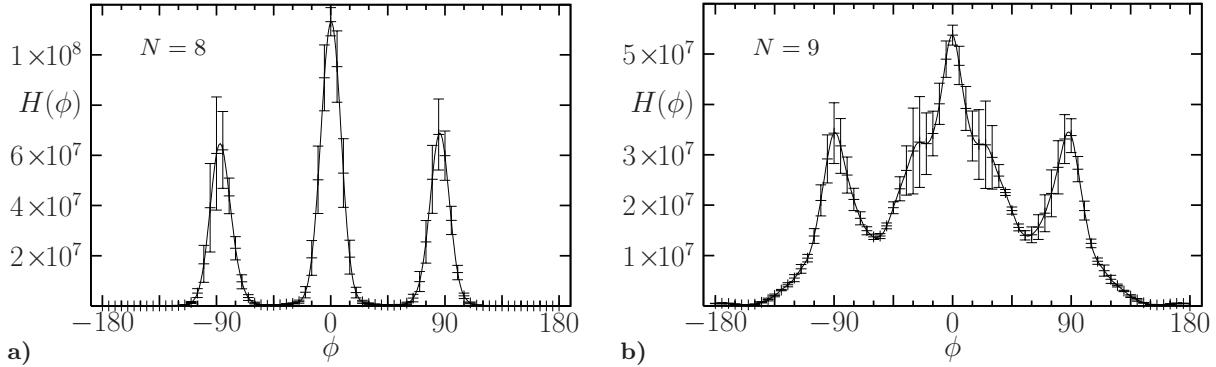}
\caption{Torsion angle distributions at $T=0.1$ for the \textbf{a})
$N=8$ and
\textbf{b}) $N=9$ polymers in phase $\alpha$ at $\rho=0.7$
and $\rho=0.72$, respectively (cuboid or sc-helical
region). Each histogram contains about $10^{10}$ entries. For visualizations of corresponding
conformations see, e.g., Fig.~\ref{fig1}\,c).}
\label{fig8II}
\end{figure*}
\begin{figure*}
\includegraphics{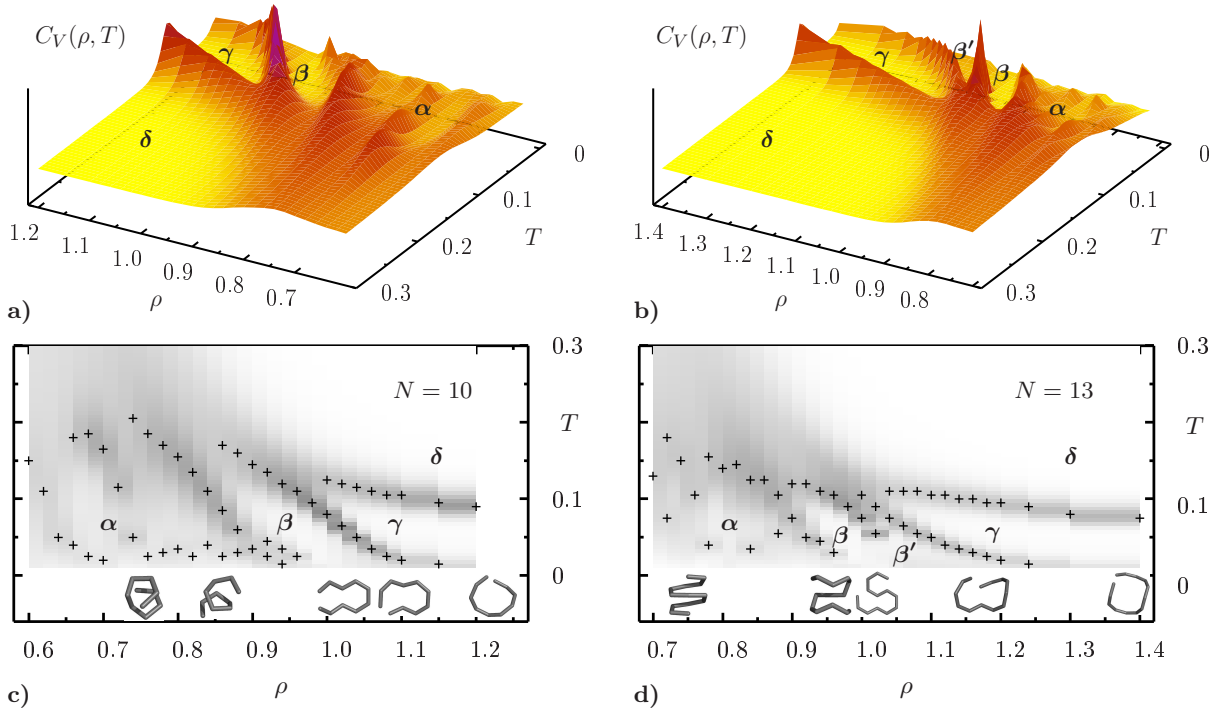}
\caption{(color online) Phase diagrams of the $N=10$ (left) and
$N=13$ (right) polymers analogously to Fig.~\ref{fig1}.}
\label{fig7}
\end{figure*}
The region of lowest thickness, $\alpha$, is the helical
phase. This phase can be further separated into subphases,
where in one of them the exact $\alpha$-helix resides as a
ground state for $N=8$ and $N=9$~\cite{partI}. In a further
region, simple-cubic helical structures~\cite{dill90jcp}, or
cuboids for $N=8$, corresponding to the ground-state
conformations in the range
$1/\sqrt{2}\approx0.707\leq\rho\lesssim0.8$, respectively,
dominate~\cite{foot1}. These regions are separated by
noticeable, but in the context of the whole phase diagram
less important, transition lines. For illustration, we show
in Fig.~\ref{fig8II} the distribution of torsional angles in
the cuboid region for $N=8$, $\rho=0.7$ and $N=9$,
$\rho=0.72$ at temperature $T=0.1$. For the $N=8$ polymer,
it can clearly be seen that only conformations with
torsional angles of $0$ and $\pm\pi/2$, i.e., cuboids,
occur. For the $N=9$ polymer, these angles are still
dominant, although not occurring exclusively. In any case,
the existence of that region is insofar worth mentioning as
the corresponding conformations do not appear as ground
states for this length and as it shows that it is a
characteristic feature and not only a length-dependent
artefact.

Figure~\ref{fig7} shows the phase diagrams for the longer
tubes consisting of $N=10$ and $N=13$ monomers analogously
to Fig.~\ref{fig1}. In general, beside the short-length
artefacts near $T=0$, the phase diagrams at different
lengths do not differ qualitatively much from each
other. The general thermodynamic behavior is quite similar
for all system sizes, especially we find again the four
major phases discussed above. Also, the characteristics of
the sprawled-coil and bended-ring regions do not depend,
beside an obvious shift of the thickness parameter, on the
polymer length. We note, however, the onset of the formation
of tertiary structures, as also discussed in
Ref.~\cite{partI}, especially the helical phase $\alpha$
becomes internally more complex. Furthermore, the relevant
thermodynamical activity shifts to lower \hbox{temperatures}.

The ground-state conformations for these systems, plotted
again in the insets of Fig.~\ref{fig7}\,c) and~d), support
our interpretation of the phases given above. Especially the
motivation for denoting $\beta$ the sheet phase becomes
clearer, as we found almost planar, ``two-dimensional''
ground states seeming to crystallize on a honeycomb
lattice. These conformations are the dominant conformations
in $\beta$ at finite temperatures as well and form, in the
case of the $N=13$ polymer, three LJ contacts, in the sense
of a contact map~\cite{partI}. We find a further interesting
detail here, which occurs only for these longer chains. The
13mer is long enough, that an intermediate phase
$\beta^\prime$ emerges between $\beta$ and $\gamma$. This
phase is populated, as indicated by the ground-state
conformation shown in Fig.~\ref{fig7}\,d), by two small bended
circles such that two LJ contacts are formed.

Since we focus in our study on the very precise investigation of short chains 
only in order to elaborate the thickness and temperature
dependence of secondary-structure formation, noticeable
tertiary effects, such as the globular arrangement of
secondary-structure segments, are not yet relevant. For
longer chains, a classification of structural phases is only
possible by accounting for the globular tertiary folding
behavior as it was shown in Ref.~\cite{hoang1}, where
protein-like structures were identified as marginally
compact, thus representing a particular globular
conformational phase.

\section{\label{sec:AB}Secondary-structure pseudophases of a 
hydrophobic-polar tube model}

As the central result of this work, we have shown above how
the sole introduction of a thickness constraint enhances the
formation of different secondary structures, including helix
and sheet formation, for classes of homopolymers. Here, we
modify the homopolymer tube model by introducing two species
of monomers: hydrophobic (A) and hydrophilic or polar (B)
ones. The nonbonded Lennard-Jones interaction between pairs
of monomers now depends on their types:
\begin{equation}
V_{\text{LJ}}^{\text{AB}}(r_{i,j})=4\left(\frac{1}{r_{i,j}^{12}}-\frac{C(i,j)}{r_{i,j}^6}\right)\,,
\end{equation}
where
\begin{equation*}
C(i,j)=\begin{cases}
+1&\text{for AA contacts}\,,\\
+1/2&\text{for BB contacts}\,,\\
-1/2&\text{for AB contacts}\,.
\end{cases}
\end{equation*}
Besides the strong attraction of A-type monomers we thus
have a weak attraction between B-type monomers and a weak
repulsion between monomers of different type, favoring
``hydrophobic'' core formation of A monomers. To enable a
direct comparison with the literature on the standard
linelike AB
model~\cite{still93pre,still95pre,hsugrass03pre,arkin05pre},
we introduce here in addition a bending term and take the
total energy as
\begin{equation}
E_{\text{AB}}(\textbf{X})=\frac{1}{4}\sum_k\left(1-\cos
\vartheta_k\right)+
\sum_{i,j=i+2}V_{\text{LJ}}^{\text{AB}}(r_{i,j})\,,
\end{equation}
where the $\vartheta_k$'s are the bending angles of adjacent
bond vectors.

\begin{figure*}
\includegraphics{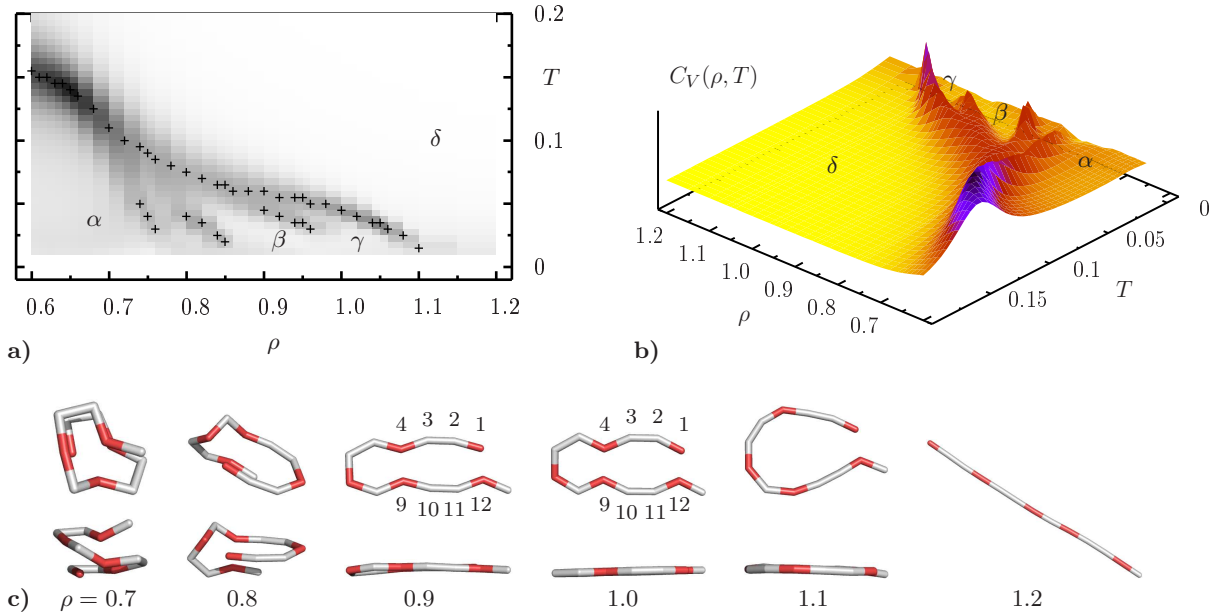}
\caption{(color online) Pseudophase diagram of the $N=13$ Fibonacci AB
heteropolymer. \textbf{a)} The plot shows the top-view with
marked peak positions of the specific heat for various
parameters $\rho$, \textbf{b)} the qualitative view of the
specific-heat landscape. Gray scales encode the value of the
specific heat. The pictures in
\textbf{c)} illustrate selected ground-state
conformations. Conformations are shown from different
viewpoints, A monomers are marked by red color (dark gray),
B monomers are white.}
\label{figAB}
\end{figure*}
Just to acquire a taste for the effects of these changes, we
show as an example results for the 13mer Fibonacci sequence
AB$_2$AB$_2$ABAB$_2$AB, which has been studied in the
linelike AB model, i.e., with $\rho=0$, some time
ago~\cite{still95pre,hsugrass03pre,arkin05pre}.
Figure~\ref{figAB} shows the phase diagram analogously to
Figs.~\ref{fig1} and~\ref{fig7}, as well as selected
ground-state conformations. The general structure including
several separated structural subphases is similar to that
for the presented homopolymers. The most prominent finding
is definitely the very stable $\beta$-sheet region in the
interval $0.90\leq\rho\leq1.01$, as $T\to0$. The
conformations there are neither of $\kappa_0$- nor
$\tau_0$-type, i.e., they have neither constant bond nor
torsion angles~\cite{partI}, but they are indeed ``planar''
(data not shown, see Fig.~\ref{figAB} for
visualization). These qualitative properties do not change
over the entire region.  A~quantitatively remarkable fact is
the variation of the intra-monomer distances. We note, that
the interaction length between the opposite hydrophobic A
monomers $1-12$ ($r_{1,12}=1.13$, see Fig.~\ref{figAB}\,c)
for monomer numbering) and $4-9$ ($r_{4,9}=1.15$) in the
sheet conformation does not change in the whole thickness
region at all. On the other hand, the distances between the
B monomers $2-11$ and $3-10$ increase ($\Delta
r_{2,9}=\Delta r_{3,10}=0.27$) and decrease between the A
monomers $1-4$ and $9-12$ ($\Delta r_{1,4}=\Delta
r_{9,12}=-0.10$, differences respecting the conformations at
$\rho=0.9$ and $\rho=1.0$). The van-der-Waals attraction
between the A monomers is thus the dominant factor that
stabilizes the $\beta$-sheet. Remarkably, as becomes clear
by the listed geometrical quantities above, the bending
energy is even increasing with increasing thickness in this
region, contrarily to the general overall trend, that the
bending energy decreases with increasing thickness. We
discuss the influence of the bending term further
below. Remember that there are planar six-ring conformations
at comparable thicknesses for the $N=8$, $N=10$, and $N=13$
homopolymer ground states~\cite{partI}. These structures are
now stabilized by the specific monomer sequence. We
emphasize that the tube thickness keeps playing an important
role. Just simulating the given sequence in a
two-dimensional space without thickness leads to completely
different conformations, consisting of a hydrophobic core
and a polar shell~\cite{still95pre}.

At lower thickness parameters we find structures with
helical properties, which, however, depend on the monomer
sequence. We note here a very pronounced conformational
transition from random coils to native conformations at
$0.1\lesssim T\lesssim0.15$, which is in detail discussed
for the linelike limit ``$\rho\lesssim0.6$'' in
Ref.~\cite{arkin05pre}. With increasing thickness the
ground-state conformation becomes a ring and finally
switches to a stretched rod, which, contrarily to the
homopolymers discussed above, appears as ground-state
conformation.
This is a qualitative difference to the results in
Sec.~\ref{sec:results_homo}.

Finally two remarks are in order. Firstly, using the
described model, we make two independent changes compared to
the homopolymer model used before. We introduce different
kinds of monomers with different interactions and in
addition a bending stiffness. To evaluate the influence of
each of the two changes, we simulated the 13mer with a
homopolymer sequence consisting of just hydrophobic A
monomers (A$_{13}$), which is equal to the homopolymer
studied without bending stiffness in
Sect.~\ref{sec:results_homo}. We made sure, that the
influence of the bending stiffness is marginal for both,
ground-state structures and thermodynamical behavior in the
relevant structural regions. The ground-state energies
change by 1\% to 5\% in the $\alpha$ and $\beta$ region,
the structures themselves remain qualitatively the same. The
effect on the thermodynamical behavior is marginal, in
particular peak positions in the specific heat are not
influenced. We conclude, therefore, that the described
behavior is predominantly based on the influence of
different monomer types. Remember also the example discussed
above on this observation. Note that choosing a B
homopolymer (B$_{13}$) would correspond to $\sigma=2^{1/6}$ and
$\epsilon=1/4$ in Eq.~(\ref{eq:LJ}), with
$r^\text{opt}_{i,j}=2^{1/3}$ and
$V_\text{LJ}(r^\text{opt}_{i,j})=-1/4$. Absorbing the energy
scale in the definition of temperature (i.e.,
$\epsilon=1/4\to\epsilon_\text{B}=1$), we would work with
$T_\text{B}=T_\text{A}/4$.

Secondly, as a methodological remark, knowing that ground
states of one-dimensional linelike models do intrinsically
have some measurable ``natural thickness'' $d(\mathbf{X})$
in the meaning of the interpretation of the global radius of
curvature, see Eq.~(\ref{eq:grc}), it may be favorable to
search for ground states by simulating the polymer with a
thickness constraint slightly below this value. One
restricts the conformational space significantly and may
travel much faster through the remaining phase
space. That way, we could confirm for the 13mer
Fibonacci sequence and other widely-used AB polymers with
$N\leq21$ monomers the ground-state energies and conformations
presented over the past
years~\cite{arkin05pre,kim05pre,elser06pre,huang06pre}.

\section{\label{sec:summary}Summary}

We present in this article results of a computer simulation
study of the thermodynamical behavior of a tube model for
simple homopolymers as well as for an exemplified
hydrophobic-polar heteropolymer. The thickness of the tube
in our simulations is controlled by a single parameter, the
global radius of curvature, which depends on three-body
interactions~\cite{foot2}.

After focusing on ground states of homopolymers and their
properties in a previous work~\cite{partI}, we identified
dominant structural pseudophases at finite temperatures,
i.e., specific-heat landscapes depending on the thickness
parameter and temperature, representing the conformational
phase diagram. Independently of the polymer length, we find
four major structural phases. These include helices,
sheetlike planar structures, bended rings and sprawled
random coils. These different secondary structure phases can
be assigned to different ranges of the tube thickness. The
thickness parameter is therefore suitable for a
classification of the secondary structures
of polymers. Concentrating 
on the analysis of the secondary-structure formation 
of short chains, tertiary effects could widely be excluded.
Symmetries and anisotropy in the arrangement of secondary-structure
segments in globular domains~\cite{hoang1}, which are particularly interesting
for proteins, are necessarily of importance in the discussion 
of the folding behavior of longer chains. A precise investigation
of the thickness-dependent influence of thermal fluctuations on
the phase structure is future work.

In an extension of the tube polymer concept, we also introduced the AB tube model for
hydrophobic-polar heteropolymers and discussed results for a
given sequence of monomers, which has extensively been
studied before without thickness. We showed that a sequence of hydrophobic and polar monomers
can stabilize the general
secondary structures. In particular we find a very
pronounced and stable region of a $\beta$-sheet structure.

Our results are qualitative in a sense that they represent the
general frame of possible conformational phases of secondary structures for thick polymers
and proteins. This is the basis of the further
analysis of pseudophases of models designed for specific
polymers or \hbox{proteins}.

To conclude, the tube picture is well suited to mimic the
volume extension of polymers, for example due to side chains
of amino acids in biopolymers. It may be employed in other
contexts as well, for example, for simulations of a tube
model for entangled networks of polymers, where the
hypothetical tube around a polymer models the suppression of
transverse undulation by the
network~\cite{degennes_book,hinsch07epjd}. Finally also the
diffusion of knots in knotted DNA can proceed via the
solitonic diffusion of compact knot
shapes~\cite{grosb07prl}. The tube picture also may be
applicable here.

\begin{acknowledgments}
This work is partially supported by the DFG (German Science
Foundation) under Grant Nos.\ JA 483/24-1/2/3 and the
Leipzig Graduate School of Excellence ``BuildMoNa''. Support
by the supercomputer time grant of the John von Neumann
Institute for Computing (NIC), Forschungszentrum J\"ulich,
is acknowledged. We thank Sebastian Sch\"obl for interesting
discussions.
\end{acknowledgments}


\begin{thebibliography}{42}
\bibitem[{foo({\natexlab{a}})}]{foot1}
\bibinfo{note}{The occurrence of these structures as ground states for certain
  system sizes, e.g., the crystallization on regular lattices as singular
  solutions of the theory, has been analyzed and discussed in detail in
  Ref.~\cite{partI}.}
\bibitem[{foo({\natexlab{b}})}]{foot2}
\bibinfo{note}{For the advantages of this implementation and differences to
  realizations with pairwise repulsive potentials, we refer the reader to the
  discussions in Refs.~\cite{banagonz03jsp,letter}.}
\end{thebibliography}
\end{document}